\documentclass{article}

\usepackage{arxiv}

\usepackage[utf8]{inputenc} 
\usepackage[T1]{fontenc}    
\usepackage[pagebackref=true]{hyperref}       
\usepackage{url}            
\usepackage{booktabs}       
\usepackage{amsfonts}       
\usepackage{nicefrac}       
\usepackage{microtype}      
\usepackage{lipsum}		    
\usepackage{graphicx}
\usepackage{natbib}
\usepackage{doi}
\usepackage{tikz-cd}

\usepackage{amsmath}
\usepackage{amscd}
\usepackage{amsthm}
\usepackage{amssymb}

\usepackage{booktabs}
\usepackage{listings}

\usepackage{xcolor}

\definecolor{cqlkw}{RGB}{0,0,180} 
\definecolor{cqlcmt}{RGB}{0,128,0} 
\definecolor{cqltype}{RGB}{128,128,128} 
\definecolor{cqlstr}{RGB}{163,21,21} 

\lstdefinelanguage{CQL}{
  keywords={ 
    literal, external_types, external_parsers, external_functions, entities, foreign_keys, attributes, path_equations, entity_equations,
    generators, equations,  quotient, getSchema, getMapping, 
    sigma, delta, chase, coproduct, eval, simple, forall, where, from, options},
  keywordstyle=\color{cqlkw}\bfseries,
  sensitive=true,
  morecomment=[l]{\#},
  commentstyle=\color{cqlcmt},
  morestring=[b]",
  stringstyle=\color{cqlstr},
  classoffset=1,
  morekeywords={typeside, schema, instance, schema_colimit,mapping, constraints, query},
   keywordstyle=\color{cqlstr}\bfseries,
   classoffset=0,
}

\lstdefinestyle{cql}{
  language=CQL,
  basicstyle=\ttfamily\small,
  frame=single,
  framerule=0.4pt,
  rulecolor=\color{gray!60},
  backgroundcolor=\color{gray!5},
  numbers=left,
  numberstyle=\tiny\color{gray},
  numbersep=6pt,
  tabsize=4,
  showstringspaces=false,
  breaklines=true,
  columns=flexible,
  xleftmargin=12pt,
  aboveskip=8pt,
  belowskip=8pt,
  captionpos=b,
}

\newtheoremstyle{defstyle}
  {\topsep}   
  {\topsep}   
  {\normalfont}  
  {}          
  {\bfseries} 
  {.}         
  {.5em}      
  {}          

\theoremstyle{defstyle}
\newtheorem{innerdefinition}{Definition}[section]

\theoremstyle{defstyle}

\usepackage{todonotes}

\title{A Categorical Approach to Semantic Interoperability across Building Lifecycle}

\author{ \href{https://orcid.org/0000-0002-6014-3228}{\includegraphics[scale=0.06]{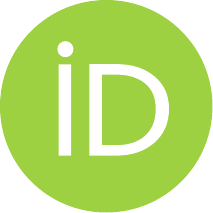}\hspace{1mm}Zoltan Nagy}\\
	Department of Built Environment\\
	Eindhoven University of Technology\\
	Eindhoven, The Netherlands \\
	\texttt{z.nagy@tue.nl} \\
	\And
    \href{https://orcid.org/0000-0002-4768-7861}{\includegraphics[scale=0.06]{orcid.pdf}\hspace{1mm}Ryan Wisnesky} \\
	Conexus AI, Inc.\\
	San Francisco,CA, USA \\
	\texttt{ryan@conexus.com} \\
    \And
    \href{https://orcid.org/0000-0000-0000-0000}{\includegraphics[scale=0.06]{orcid.pdf}\hspace{1mm}Kevin Carlson} \\
	Topos Institute\\
	Berkeley, CA, USA \\
	\texttt{kevin@topos.institute} \\
    \And
	\href{https://orcid.org/0000-0002-4639-627X}{\includegraphics[scale=0.06]{orcid.pdf}\hspace{1mm}Eswaran Subrahmanian} \\
	Carnegie Mellon University\\
	Pittsburgh, PA, USA \\
	\texttt{es3e@andrew.cmu.edu} \\
	\And
	\href{https://orcid.org/0000-0002-6601-2237}{\includegraphics[scale=0.06]{orcid.pdf}\hspace{1mm}Gioele Zardini} \\
	Laboratory for Information and Decision Systems\\
    Massachusetts Institute of Technology\\
	Cambridge, MA, USA \\
	\texttt{gzardini@mit.edu} \\
}

\renewcommand{\shorttitle}{A Categorical Approach for Building Data Integration}

\begin{document}
\maketitle

\begin{abstract}
Buildings generate heterogeneous data across their lifecycle, yet integrating these data remains a critical unsolved challenge. Despite three decades of standardization efforts, over 40 metadata schemas now span the building lifecycle, with fragmentation accelerating rather than resolving. Current approaches rely on point-to-point mappings that scale quadratically with the number of schemas, or universal ontologies that become unwieldy monoliths. The fundamental gap is the absence of mathematical foundations for structure-preserving transformations across heterogeneous building data. Here we show that category theory provides these foundations, enabling systematic data integration with $O(n)$ specification complexity for $n$ ontologies. We formalize building ontologies as first-order theories and demonstrate two proof-of-concept implementations in Categorical Query Language (CQL): 1) generating BRICK models from IFC design data at commissioning, and 2) three-way integration of IFC, BRICK, and RealEstateCore where only two explicit mappings yield the third automatically through categorical composition. Our correct-by-construction approach treats property sets as first-class schema entities and provides automated bidirectional migrations, and enables cross-ontology queries. These results establish feasibility of categorical methods for building data integration and suggest a path toward an \emph{app ecosystem} for buildings, where mathematical foundations enable reliable component integration analogous to smartphone platforms.

\end{abstract}
\section{Introduction}

Despite 30 years of standardization efforts, semantic interoperability remains the primary barrier to effective building data integration. In fact, nearly a decade after a comprehensive review~\cite{pauwels_semantic_2017}, recent analysis observes that construction informatics remains \emph{``fragmented and lacks clarity in understanding the interconnection of different data management strategies"}~\cite{bucher_bim_2024}. This fragmentation is accelerating rather than resolving: over 40 metadata schemas have been identified spanning the building lifecycle, with 60\% developed within the previous five years, i.e., there is a divergence toward competing standards rather than convergence toward shared solutions~\cite{pritoni_metadata_2021}. 

In terms of applications, the authors of \cite{pritoni_metadata_2021} also evaluated five major building ontologies (BOT, SSN/SOSA, SAREF, RealEstateCore, BRICK) against three common use cases: energy audits, automated fault detection, and optimal control. They concluded that none fully supported all three use cases because building modelers find it ``difficult, labor intensive, and costly to create, sustain, and use semantic models with existing ontologies"~\cite{pritoni_metadata_2021}.  Manual, and labor-intensive processes requiring expert knowledge is also considered one of the main reasons that the ``BRICK Schema has not been widely used in real buildings''~\cite{li_developing_2024}, and that design information captured in Industry Foundation Classes (IFC)~\cite{noauthor_industry_nodate} cannot be automatically translated to operational BRICK models~\cite{balaji_brick_2016}, leading to information loss when a building transitions into operation~\cite{lange_evaluation_2018}.

The fundamental challenge is that these data exist in incompatible formats using different semantic models, creating data silos that resist integration. Of course, each representation serves its intended purpose effectively within its domain. Yet, their structural and semantic differences create fundamental barriers to integration.

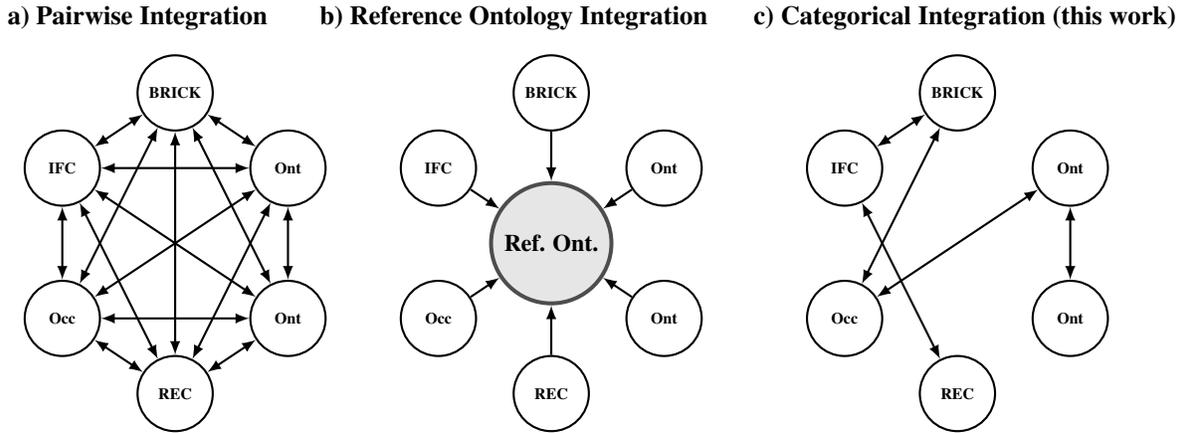
\begin{figure}[tbh]
\centering
\begin{tikzpicture}[
    node distance=1.5cm and 1.5cm,
    schema/.style={
        circle,
        minimum width=1cm,
        text centered,
        draw=black,
        thick,
        font=\tiny\bfseries,
        fill=white
    },
    colimit/.style={
        circle,
        minimum width=1.6cm,
        text centered,
        draw=black!70,
        line width=1.5pt,
        fill=black!10,
        font=\small\bfseries
    },
    pairwise/.style={
        <->,
        >=latex,
        thick,
        draw=black
    },
    functor/.style={
        <->,
        >=latex,
        thick,
        draw=black
    },
    mapping/.style={
        ->,
        >=latex,
        thick,
        draw=black
    }
]

\node[font=\bfseries] at (-5.5, 3.5) {a) Pairwise Integration};

\node[schema] (t1) at (-6.5, 1.5) {IFC};
\node[schema] (t2) at (-5, 2.5) {BRICK};
\node[schema] (t4) at (-6.5, -0.5) {Occ};
\node[schema] (t5) at (-5, -1.5) {REC};
\node[schema] (t6) at (-3.5, -0.5) {Ont};
\node[schema] (t7) at (-3.5, 1.5) {Ont};

\draw[pairwise] (t1) -- (t2);
\draw[pairwise] (t1) -- (t4);
\draw[pairwise] (t1) -- (t5);
\draw[pairwise] (t2) -- (t4);
\draw[pairwise] (t2) -- (t5);
\draw[pairwise] (t4) -- (t5);

\draw[pairwise] (t1) -- (t6);
\draw[pairwise] (t1) -- (t7);
\draw[pairwise] (t2) -- (t6);
\draw[pairwise] (t2) -- (t7);
\draw[pairwise] (t4) -- (t6);
\draw[pairwise] (t4) -- (t7);
\draw[pairwise] (t5) -- (t6);
\draw[pairwise] (t5) -- (t7);
\draw[pairwise] (t6) -- (t7);

\node[font=\bfseries] at (-0.5, 3.5) {b) Reference Ontology Integration};

\node[colimit] (U) at (0, 0.5) {Ref. Ont.};

\node[schema] (u1) at (0, 2.5) {BRICK};
\node[schema] (u4) at (0, -1.5) {REC};
\node[schema] (u5) at (-1.5, -0.5) {Occ};
\node[schema] (u6) at (-1.5, 1.5) {IFC};
\node[schema] (u7) at (1.5, -0.5) {Ont};
\node[schema] (u8) at (1.5, 1.5) {Ont};

\draw[mapping] (u1) -- (U);
\draw[mapping] (u4) -- (U);
\draw[mapping] (u5) -- (U);
\draw[mapping] (u6) -- (U);
\draw[mapping] (u7) -- (U);
\draw[mapping] (u8) -- (U);

\node[font=\bfseries] at (5.5, 3.5) {c) Categorical Integration (this work)};

\node[schema] (c1) at (5.4, 2.5) {BRICK};
\node[schema] (c4) at (5.4, -1.5) {REC};
\node[schema] (c5) at (3.9, -0.5) {Occ};
\node[schema] (c6) at (3.9, 1.5) {IFC};
\node[schema] (c7) at (6.9, 1.5) {Ont};
\node[schema] (c8) at (6.9, -0.5) {Ont};

\draw[functor] (c1) -- (c6);
\draw[functor] (c1) -- (c5);
\draw[functor] (c6) -- (c4);
\draw[functor] (c5) -- (c7);
\draw[functor] (c7) -- (c8);

\end{tikzpicture}
\caption{Comparison of pairwise, reference, and categorical integration of $n$ ontologies for building digital twins. a) Pairwise integration requires $O(n^2)$  mappings with no composition guarantees (order matters). b) Reference ontology requires $O(n)$ mappings into a predetermined universal schema, but transformations are one-way and lossy. c) Categorical integration requires $O(n)$ mappings to connect all schemas bidirectionally, enabling data exchange between any ontology pair through (path independent) composition -- the best of both previous approaches. \textbf{Abbreviations} (as example): IFC = Industry Foundation Classes, BRICK, Occ = Occupant behavior, REC = RealEstateCore, Ont = generic ontologies}
\label{fig:digital_twin_architecture}
\end{figure}

Generally, data integration approaches fall into two categories, each with significant limitations~\cite{bucher_bim_2024}: 1) Point-to-point mappings between specific schemata, which become quadratically (or even exponentially, when point order matters) complex, and require custom translations that must be manually maintained as schemata evolve ((Fig.~\ref{fig:digital_twin_architecture}a);); and 2) Reference ontologies that attempt to map all schemata to a single universal ontology, e.g., BRICK, but struggle with evolving domain requirements, and often become unwieldy monoliths that satisfy no stakeholder completely~\cite{schultz_algebraic_2017} ((Fig.~\ref{fig:digital_twin_architecture}b);). Linked-data technologies (RDF, SPARQL, OWL) provide infrastructure for both approaches but do not themselves resolve these fundamental scaling challenges: implementations still require either $O(n^2)$ pairwise alignments for $n$ ontologies or commitment to a predetermined (and single) reference ontology. In contrast, as we will see in this work, categorical integration (Fig.~\ref{fig:digital_twin_architecture}c) requires only $O(n)$ specifications while enabling bidirectional data exchange between any pair of ontologies, requiring much less work and being much more adaptable than both approaches.

This persistent integration and fragmentation challenge hints at a \textbf{fundamental gap: the absence of mathematical foundations} for structure-preserving transformations across heterogeneous building data. As buildings are inherently multi-scale, multi-physics, multi-stakeholder systems where different representations serve different purposes, semantic interoperability frameworks must preserve their individual strengths without becoming prohibitively complex to manage in practice.

Category theory provides the mathematical foundations that building data integration needs. Developed since the 1940s to connect different areas of mathematics, category theory studies structural relationships independent of implementation details. Its fundamental viewpoint is that \textbf{many properties of systems can be understood by studying relationships between objects rather than the objects' internal details}. Category theory studies the preservation and transformation of relationships using functorial (homomorphic) mappings, which is crucial to building data integration, because building data integration is fundamentally about preserving relationships during transformation. The critical insight is that the way ontologies such as BRICK and IFC compose and relate can be studied by category theory, once they have been axiomatized in first order logic, as we will see shortly.

Category theory has proven transformative across industries: enabling consensus-free integration of formulae-laden excel spreadsheets in oil and gas without combinatorial explosion~\cite{baylor_consensus-free_2025}, streamlining schema integration in manufacturing and databases~\cite{wisnesky_using_2017, islam_categorical_1994}, demonstrating 10$\times$ to 100$\times$ speedups in certain big data algorithms~\cite{thiry_categories_2018}, and enabling compositional co-design of autonomous systems~\cite{zardini_phdthesis, censi2024}. Beyond efficiency, categorical methods enforce data constraints~\cite{johnson_implementing_2008}, guarantee algebraic interoperability~\cite{schultz_algebraic_2017}, and support scalable parallel processing~\cite{koupil_unified_2022}. Category theory provides ``more flexible and robust integration of existing and evolving information'', whereas the reference (universal) ontology approach requires ``characterizing entire portal databases in terms of additional information''~\cite{wisnesky_using_2017}; that is, the reference ontology must be an ontology at least as expressive as every ontology mapping into it, a difficult condition to guarantee; in our approach, the reference ontology is automatically synthesized from the $O(n)$ mappings provided.  (In fact, our approach will synthesize a different reference ontology for different input mappings, allowing users to exchange data in more than one way).  Thus, category theory provides correct-by-construction algorithms for data migration between domain models, data aggregation and model integration, which are exactly the capabilities needed for building lifecycle data integration.

In this paper, we demonstrate how categorical methods apply to building data integration through two proof-of-concept implementations. We formalize simplified IFC, BRICK and REC schemas using first-order logic, and use the open-source Categorical Query Language (CQL, available at \url{http://categoricaldata.net}) to automate data exchange between them. Our contributions are:

\begin{enumerate}
\item \textbf{First-order formalization of building ontologies:} We propose to represent building ontologies like IFC, BRICK, etc., as theories in first-order logic, making explicit the mathematical structure underlying building data models.

\item \textbf{Integration through theory extensions:} We demonstrate a principled way to combine $n$ incompatible ontologies by specifying how $O(n)$ formulae can define a theory extension that enables automatic data exchange across all building instances of those ontologies.  This theory extension is interpreted categorically as a ``lifting problem''~\cite{spivak_database_2014} that the CQL logic engine solves using sophisticated algorithms~\cite{meyers_fast_2022}, resulting in correct-by-construction, quality-optimal, data exchange or integration.  We also provide a user-friendly shorthand notation for specifying theory extensions.

\item \textbf{Two integration scenarios:} (1) A commissioning scenario generating BRICK operational models from IFC design data replicating a case study from literature; (2) An integration scenario demonstrating data exchange between three ontologies while only mapping two pairwise.

\end{enumerate}

With these proof-of-concept examples, we establish feasibility of categorical approaches for real building integration challenges and identify the path toward production deployment. We take an expedient approach to presenting categorical data integration: we first present the general integration pattern to build intuition in Section~\ref{s:integration_pattern}, and then, we work through the two proof-of-concept examples in Section~\ref{s:examples}.
In this way, category theory itself becomes an implementation detail rather than a prerequisite to understand this paper.
Finally, in Section~\ref{s:discussion} we discuss the implications of our work and potential future research directions, before concluding the paper in Section~\ref{s:conclusion}.

\section{Categorical data integration/exchange: A bird's eye view}
\label{s:integration_pattern}
The key insight for categorical data integration/exchange (mathematically, our framework treats integration and exchange as the same problem) is that we construct a theory extension $C$ that contains both source schemas $A$ and $B$, plus an explicit, albeit partial, specification of how $A$ and $B$ relate. Rather than writing procedural code to map data from $A$ to $B$ or vice versa, or creating an entire reference ontology that subsumes both $A$ and $B$ (a full specification of how $A$ and $B$ relate), users declare arbitrary logical relationships between $A$ and $B$ using formulae and let algorithms derive the associated reference ontology and data migration/integration, as a consequence, similar to solving a system of equations. 

The integration/migration pattern can be broken down into a design decision that has to be made once (how to (partially) relate BRICK and IFC), by a human or computer, and an execution part that is automated (how to translate data from BRICK to IFC and vice versa, or how to translate IFC and BRICK into a combined schema):

\textbf{Input specification (given once):}
\begin{itemize}
\item Which entities/tables correspond across schemas
\item Which columns/attributes correspond across corresponding tables
\item Which rows/records correspond across corresponding columns/attributes (specified using data-independent formulae)
\end{itemize}

We are being deliberately vague in writing ``correspond'', because a large fragment of first-order logic is available to define such correspondences, as we will see later.

\textbf{Automated (and correct-by-construction) execution:}
\begin{itemize}
\item Verifies consistency of data against schema formulae and type safety at compile time
\item Migrates data from source schemas into a synthetic combined schema (data integration)
\item Migrates data from the combined schema back to source schemas (data exchange)
\item Round-trips data from each source through the combined schema and back to each source, quantifying information gain or loss on a table-by-table basis (not described in this paper)
\end{itemize}

We will use (functional) first-order logic as the syntax for mappings/formulae/ontologies/etc in this paper.  In first-order logic, a {\it signature} consists of a set of {\it type} names, and a set of typed function symbols.  A {\it theory} over a signature is a set of formulae that describe the intended meaning of the symbols in the signature.  

For example, the ``signature of arithmetic'' is defined using a single type, $Nat$ (natural number), and four typed function symbols, $0 : Nat$ and $S : Nat \to Nat$ (successor, so that $1 = S(0)$) and $+ : Nat \to Nat$ and $\times : Nat \to Nat$, and the ``theory of arithmetic'' is the formulae: 
$$
\forall x:Nat, \ x + 0 = x \ \ \ \ \ \ \ \ \  \ \forall x:Nat, y:Nat, \ S(x) + y = S(x+y) \ \ \ \ \ \ \ \ 
$$
$$
\forall x:Nat, \ x \times 0 = 0 \ \ \ \ \ \ \ \ \forall x:Nat,\  x \times 1 = 1 
\ \ \ \ \ \ \ \ \ \forall x:Nat, y:Nat, \ S(x) \times y = y + (x \times y)
$$
A {\it model} $I$ of a theory $T$ over a signature $S$ provides a set $I(s)$ for each type $s \in S$, and provides, for each function symbol $f : c \to d \in S$, a function $I(f) : I(c) \to I(d)$ satisfying the formulae of $T$.  For example, we can take $I(Nat)= 0,1,2,\ldots$ and use the usual addition and multiplication functions as a model.  But we can also take e.g. $I(Nat) = true, false$ and interpret $+$ and $\times$ as logical and and logical or.  The use of first-order logic in data integration has a long history~\cite{arenas_foundations_2014}; and more recently, categorical methods decompose data exchange into finer primitives than logic-based formalisms: adjoint functors ($\Sigma$, $\Delta$) for data migration and colimits for schema composition~\cite{schultz_algebraic_2017}. This decomposition identifies theory extensions as the proper mechanism for composing ontologies.

In this paper's framework, a \textbf{schema} $S$ is a first-order theory decomposed in the following components~\cite{schultz_algebraic_2017} (not all first-order theories can be decomposed into schemas, but all schemas compose into first-order theories):

\begin{description}
\item[Type-side:] Data types and library functions (e.g., \texttt{String}, \texttt{+}, \texttt{concat})
\item[Entities:] Data tables or record types (e.g., \texttt{Equipment}, \texttt{Location}, \texttt{Point})
\item[Foreign keys:] Functions between entities (e.g., \texttt{hasLocation : Equipment → Location})
\item[Attributes:] Functions from entities to types (e.g., \texttt{name : Equipment → String})
\item[Formulae:] Constraints that instances must satisfy (e.g., $\forall e:\texttt{Equipment}, \ e.\texttt{hasLocation}.\texttt{building} = e$).  These formulae must be ``existential Horn clauses''~\cite{arenas_foundations_2014}, that is, formulae of the form $\forall x:X \ y:Y \ldots ,\  \phi(x, y, \ldots) \  \to \ \exists u:U, \ v:V, \ \ldots, \  \psi(x,y,u,v)$. 
\end{description}

An \textbf{instance} $I$ on schema $S$ is a model of $S$; that is, it assigns concrete data: e.g., a set $I(\texttt{Equipment})$ of equipment records, a function $I(\texttt{hasLocation})$ mapping each equipment to its location, values for attributes, etc.  Being models, instances must satisfy all schema formulae in the usual way~\cite{arenas_foundations_2014}. 

It can be proved that the restrictions above are the minimal required for any data exchange framework, including the one we use in this paper, to have properties desirable for data exchange/integration.  For example, if we allow arbitrary first-order formulae (or arbitrary formulae in RDF/OWL's ``description logic''), then mathematically there may be more than one solution to a data exchange problem. 

\section{Examples}
\label{s:examples}
Having established intuition for writing schemas, we now demonstrate our application through two proof-of-concept examples:

\begin{description}
\item[Example 1:] Generating BRICK operational models from IFC design data at commissioning---replicating the scenario from~\cite{mavrokapnidis_linked-data_2021} using categorical methods.
\item[Example 2:] Three-way integration of IFC, BRICK, and RealEstateCore (REC), demonstrating how categorical composition enables cross-ontology queries with only linearly-many pairwise specifications.
\end{description}

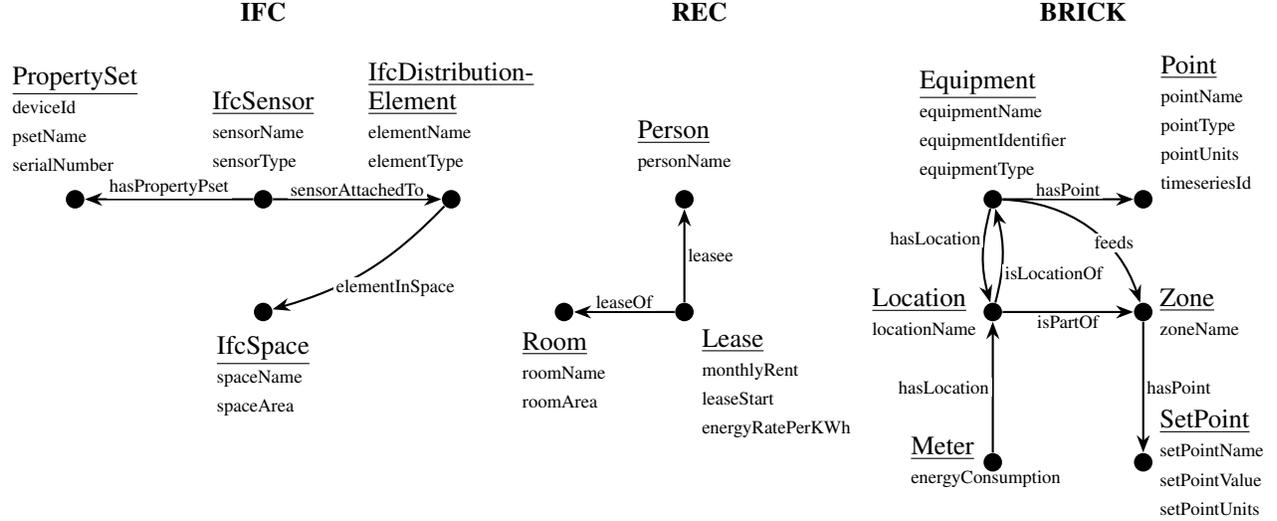
\begin{figure}[tbh]
\centering
\begin{tikzpicture}[
  node distance=1.2cm and 1.5cm,
  entity/.style={
    circle,
    fill=black,
    minimum size=7pt,
    inner sep=0pt
  },
  attribute box/.style={
    rectangle,
    align=left,
    inner sep=1pt
  },
  relationship/.style={
    ->,
    >=Stealth,
    thick
  },
  relationship label/.style={
    font=\scriptsize,
    fill=white,
    inner sep=1pt
  }
]

\begin{scope}[xshift=-6cm]

\node at (1, 4.5) {\textbf{IFC}};

\node[entity] (ifc_pset) at (-1.5, 2) {};
\node[entity] (ifc_sensor) at (1, 2) {};
\node[entity] (ifc_elem) at (3.5, 2) {};
\node[entity] (ifc_space) at (1, 0.5) {};

\node[attribute box, above=0.2cm of ifc_pset, anchor=south] {
  \underline{PropertySet}\\
  {\scriptsize deviceId}\\
  {\scriptsize psetName}\\
  {\scriptsize serialNumber}
};

\node[attribute box, above=0.2cm of ifc_sensor, anchor=south] {
  \underline{IfcSensor}\\
  {\scriptsize sensorName}\\
  {\scriptsize sensorType}
};

\node[attribute box, above=0.2cm of ifc_elem, anchor=south] {
  \underline{IfcDistribution-}\\
  \underline{Element}\\
  {\scriptsize elementName}\\
  {\scriptsize elementType}
};

\node[attribute box, below=0.15cm of ifc_space, anchor=north] {
  \underline{IfcSpace}\\
  {\scriptsize spaceName}\\
  {\scriptsize spaceArea}
};

\draw[relationship] (ifc_sensor) -- node[relationship label, above] {hasPropertyPset} (ifc_pset);
\draw[relationship] (ifc_sensor) -- node[relationship label, above] {sensorAttachedTo} (ifc_elem);
\draw[relationship] (ifc_elem) to[bend left=15] node[relationship label, right, pos=0.7] {elementInSpace} (ifc_space);
\end{scope}

\begin{scope}[xshift=5.5cm]
\node at (0.4, 4.5) {\textbf{BRICK}};

\node[entity] (brick_equip) at (-0.8, 2) {};
\node[entity] (brick_point) at (1.2, 2) {};
\node[entity] (brick_loc) at (-0.8, 0.5) {};
\node[entity] (brick_zone) at (1.2, 0.5) {};
\node[entity] (brick_meter) at (-0.8, -1.5) {};
\node[entity] (brick_sp) at (1.2, -1.5) {};

\node[attribute box, above=0.1cm of brick_equip, anchor=south] {
  \underline{Equipment}\\
  {\scriptsize equipmentName}\\
  {\scriptsize equipmentIdentifier}\\
  {\scriptsize equipmentType}
};

\node[attribute box, above right=0cm and 0.1cm of brick_point] {
  \underline{Point}\\
  {\scriptsize pointName}\\
  {\scriptsize pointType}\\
  {\scriptsize pointUnits}\\
  {\scriptsize timeseriesId}
};

\node[attribute box, left=0.05cm of brick_loc, anchor=east] {
  \underline{Location}\\
  {\scriptsize locationName}
};

\node[attribute box, right=0.05cm of brick_zone, anchor=west] {
  \underline{Zone}\\
  {\scriptsize zoneName}
};

\node[attribute box, left=1cm of brick_meter, anchor=west] {
  \underline{Meter}\\
  {\scriptsize energyConsumption}
};

\node[attribute box, right=0.05cm of brick_sp, anchor=west] {
  \underline{SetPoint}\\
  {\scriptsize setPointName}\\
  {\scriptsize setPointValue}\\
  {\scriptsize setPointUnits}
};

\draw[relationship] (brick_equip) -- node[relationship label, above] {hasPoint} (brick_point);
\draw[relationship] (brick_equip) to[bend right=15] node[relationship label, left, pos=0.3] {hasLocation} (brick_loc);
\draw[relationship] (brick_loc) to[bend right=15] node[relationship label, right, pos=0.3] {isLocationOf} (brick_equip);
\draw[relationship] (brick_equip) to[bend left=35] node[relationship label, above, pos=0.7] {feeds} (brick_zone);
\draw[relationship] (brick_loc) -- node[relationship label, below] {isPartOf} (brick_zone);
\draw[relationship] (brick_meter) -- node[relationship label, left, pos=0.5] {hasLocation} (brick_loc);
\draw[relationship] (brick_zone) -- node[relationship label, right] {hasPoint} (brick_sp);
\end{scope}

\begin{scope}[xshift=0cm]

\node at (0.8, 4.5) {\textbf{REC}};

\node[entity] (rec_person) at (0.6, 2) {};
\node[entity] (rec_lease) at (0.6, 0.5) {};
\node[entity] (rec_room) at (-1, 0.5) {};

\node[attribute box, above=0.2cm of rec_person, anchor=south] {
  \underline{Person}\\
  {\scriptsize personName}
};

\node[attribute box, below right =0.1cm and 0.1cm of rec_lease] {
  \underline{Lease}\\
  {\scriptsize monthlyRent}\\
  {\scriptsize leaseStart}\\
  {\scriptsize energyRatePerKWh}
};

\node[attribute box, below=0.1cm of rec_room, anchor=north] {
  \underline{Room}\\
  {\scriptsize roomName}\\
  {\scriptsize roomArea}
};

\draw[relationship] (rec_lease) -- node[relationship label, right] {leasee} (rec_person);
\draw[relationship] (rec_lease) -- node[relationship label, above] {leaseOf} (rec_room);
\end{scope}

\end{tikzpicture}
\caption{Individual schemas for IFC (left), REC (middle), and BRICK (right) used in the examples}
\label{fig:individual_schemas}
\end{figure}

We chose IFC, BRICK, and REC ontologies because they represent different lifecycle phases (design, operations, property management) with real integration challenges documented in the literature. The examples are intentionally verbose to illustrate the concepts clearly; the same patterns apply to other ontology combinations. The ontologies we use in the examples are shown in Fig.~\ref{fig:individual_schemas}. We implement all examples in the open-source Categorical Query Language (CQL)~\cite{brown_categorical_2019}, with complete code and screens shots provided in Appendix~\ref{app:cql_code}.

\subsection{Example 1: Integration of Static and Dynamic Data}
This example addresses one of the most important challenges in the building lifecycle: generating operational models from design data at commissioning. We replicate the scenario from~\cite{mavrokapnidis_linked-data_2021}: given an IFC model with property sets containing BMS device identifiers, generate a BRICK operational model and transfer the device identifier to it. 

The approach in~\cite{mavrokapnidis_linked-data_2021} uses IFC (technically a subset, BOT) entities to represent spatial entities and relationships; and then manually creates BRICK instances and connects them to the IFC graph through custom alignment rules. The integration mechanism relies on manually parsing IFC property sets (\texttt{IfcPropertySet}) to extract BMS device identifiers (\texttt{deviceId}) and matching them to the manually-created BRICK timeseries identifiers (\texttt{timeseriesId}).

\begin{figure}
    \centering
    \includegraphics[width=\textwidth]{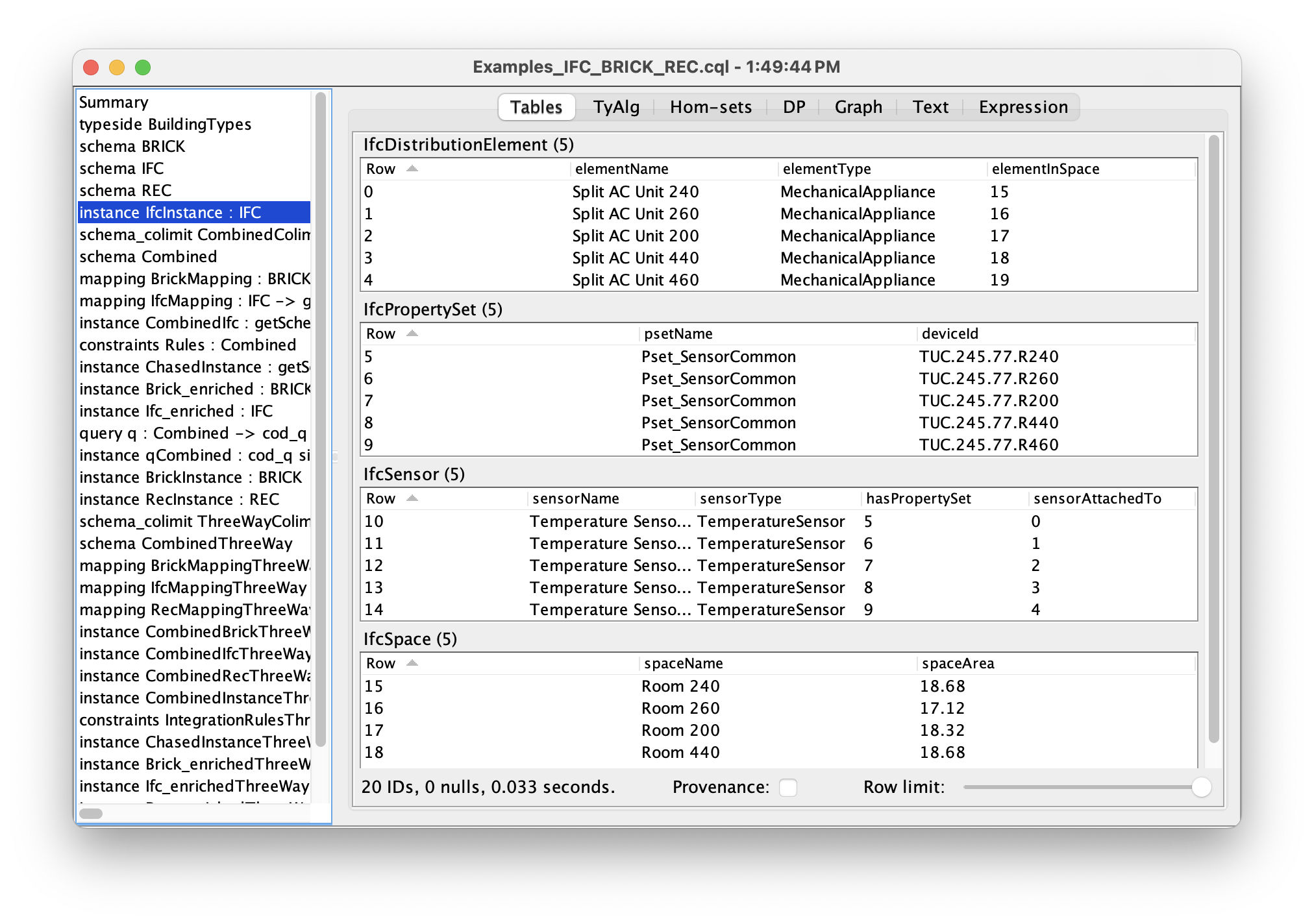}
    \caption{IFC instance for Example 1 (and 2) in CQL. The BRICK instance is initially empty (not shown).}
    \label{fig:ex1IfcInstance}
\end{figure}

We demonstrate our categorical approach on a subset of the same dataset used in~\cite{mavrokapnidis_linked-data_2021}. The example includes five rooms (named R240, R260, R200, R440, and R460) across multiple floors, each equipped with one split air conditioning unit and temperature sensor. The IFC instance (see Fig.~\ref{fig:ex1IfcInstance}) contains Spatial geometry (\texttt{IfcSpace} entities with areas), Equipment information (\texttt{IfcDistributionElement} for AC units), Temperature sensors (\texttt{IfcSensor} entities), and Property sets (\texttt{IfcPropertySet}) with BMS device tags (\texttt{deviceId}). The BRICK instance is initially empty. 

Recall that the task is to create a BRICK instance from this IFC data, which contains the \texttt{deviceId} from the IFC property set as \texttt{timeseriesId}. We begin by constructing a theory extension $C$ by specifying which tables should correspond:

\textbf{Entity Correspondences:} 
\begin{align*}
\texttt{Equipment} &\cong \texttt{IfcDistributionElement} \\
\texttt{Location} &\cong \texttt{IfcSpace}
\end{align*}

Mathematically, our theory $C$ extends BRICK and IFC with isomorphisms (inverse functions) between \texttt{Equipment} and \texttt{IfcDistributionElement} and between \texttt{Location} and \texttt{IfcSpace}, although writing these isomorphisms explicitly is verbose so we will suppress them in the first-order logic in the examples and instead identify \texttt{Equipment} with \texttt{IfcDistributionElement} and \texttt{Location} with \texttt{IfcSpace} (a technique which can itself be made precise using the notion of a ``quotient'' theory).  

\textbf{Formula 1} This equation aligns BRICK's and IFC's spatial relationship:
\begin{align*}
\texttt{Equipment.hasLocation} &= \texttt{Equipment.elementInSpace} 
\end{align*}

The above equation illustrates a short-hand for writing ``paths'' of unary functions; it abbreviates:
\begin{align*}
\forall x:\texttt{Equipment},  \ \texttt{hasLocation}(x) &= \texttt{elementInSpace}(x) 
\end{align*}

Recall that $\texttt{Equipment} \cong \texttt{IfcDistributionElement}$, and so the formulae is identifying columns from both $\texttt{Equipment}$ and $\texttt{IfcDistributionElement}$; we could have instead equivalently written
\begin{align*}
\forall x:\texttt{IfcDistributionElement},  \ \texttt{hasLocation}(x) &= \texttt{elementInSpace}(x)
\end{align*}

\textbf{Formula 2} For each pair (\texttt{IfcSensor}, \texttt{Point}) attached to the same Equipment assign \texttt{deviceId} to \texttt{timeseriesId}: 
\begin{align*}
\forall s:\texttt{IfcSensor},\ & p:\texttt{Point}\ \text{where}\ p = \texttt{s.sensorAttachedTo.hasPoint} \\ \Rightarrow & p.\texttt{timeseriesId} = 
s.\texttt{hasPropertySet.deviceId}
\end{align*}

Note that we do not need to identify \texttt{Point} with \texttt{IfcSensor}. The entity correspondences and formulae above are sufficient to transfer data from IFC to BRICK.  In addition, not shown here, we could continue completing certain BRICK entries automatically, for example \texttt{locationName} or \texttt{zoneName} if available if IFC. 

Figure~\ref{fig:threewayschema} shows the theory extension $C$; the black and outlined entities form the core IFC$\leftrightarrow$BRICK integration for this example (the blue entities will be introduced in Example~2). Table~\ref{tab:mavro_results} shows the results of a query on $C$ returning \texttt{spaceName} and \texttt{spaceArea} as stored in IFC and \texttt{timeseriesId} from BRICK.  (Note that this query could also be done in e.g. SQL on the output of CQL.)  This replicates the equivalent query and result from~\cite{mavrokapnidis_linked-data_2021}. Querying the timeseries database using SQL and \texttt{timeseriesId} to compute, e.g., the mean daily temperature, would be the same as in~\cite{mavrokapnidis_linked-data_2021}. 

\begin{figure}[tbh]
\centering
\begin{tikzpicture}[
  node distance=1.5cm and 2.5cm,
  entity/.style={
    circle,
    fill=black,
    minimum size=8pt,
    inner sep=0pt
  },
  unified entity/.style={
    circle,
    draw=black,
    line width=1.2pt,
    fill=white,
    minimum size=8pt,
    inner sep=0pt
  },
  new entity/.style={
    circle,
    draw = blue,
    fill=blue!70,
    line width = 1.2pt,
    minimum size=8pt,
    inner sep=0pt
  },
  attribute box/.style={
    rectangle,
    align=left,
    font=\footnotesize,
    inner sep=3pt
  },
  relationship/.style={
    ->,
    >=Stealth,
    thick
  },
  relationship label/.style={
    font=\scriptsize,
    fill=white,
    inner sep=1pt
  }
]

\draw[dashed, thick, rounded corners=10pt, gray!60] 
    (-7.0, 4.75) -- 
    (5, 4.75) -- 
    (5, -0.15) -- 
    (0.8, -0.15) -- 
    (0.8, -3.8) -- 
    (-1.8, -3.8) -- 
    (-1.8, -0.15) -- 
    (-7.0, -0.15) -- 
    cycle;
\node[gray] at (-5.85, 0.15) {Example 1};

\node[entity] (propertyset) at (-5.5, 2) {};
\node[entity] (sensor) at (-3, 2) {};
\node[unified entity] (element) at (-0.5, 2) {};
\node[unified entity] (space) at (-0.5, -1) {};

\node[attribute box, above=0.15cm of propertyset] {
  \underline{IFC\_PropertySet} \\[1pt]
};

\node[attribute box, above=0.15cm of sensor] {
  \underline{IFC\_IfcSensor} \\[1pt]
};

\node[attribute box, above=0.15cm of element] {
  \underline{Equipment} \\[1pt]
  {\scriptsize equipmentIdentifier} \\{\scriptsize equipmentName} \\
  {\scriptsize equipmentType} \\
  {\scriptsize elementName}\\
  {\scriptsize elementType}
};

\node[attribute box, below=0.15cm of space] {
  \underline{Location} \\[1pt]
  {\scriptsize spaceName} \\
  {\scriptsize spaceArea} \\
  {\scriptsize locationName} \\
  {\scriptsize roomName} \\
  {\scriptsize roomArea}
};

\draw[relationship] (sensor) -- node[relationship label, above] {\scriptsize hasPropertySet} (propertyset);
\draw[relationship] (sensor) -- node[relationship label, above] {\scriptsize sensorAttachedTo} (element);
\draw[relationship] (element) to[bend right=20] node[relationship label, left] {\scriptsize\begin{tabular}{r}hasLocation\\elementInSpace\end{tabular}} (space);

\node[entity] (point) at (2.5, 2) {};
\node[entity] (zone) at (2.5, 0.5) {};
\node[new entity] (meter) at (2.5, -1) {};
\node[new entity](setpoint) at (5.5,-0.5){};

\node[attribute box, right=0.15cm of point] {
  \underline{BRICK\_Point} \\[1pt]
};

\node[attribute box, right=0.15cm of zone] {
  \underline{BRICK\_Zone} \\[1pt]
};

\node[attribute box, right=0.15cm of setpoint]{
    \underline{BRICK\_SetPoint} \\[1pt]
};

\node[attribute box, right=0.15cm of meter] {
  \underline{BRICK\_Meter} \\[1pt]
};

\draw[relationship] (element) -- node[relationship label, above] {\scriptsize hasPoint} (point);
\draw[relationship] (element) to[bend right=15] node[relationship label] {\scriptsize feeds} (zone);
\draw[relationship] (space) to[bend left=15] node[relationship label] {\scriptsize isPartOf} (zone);
\draw[relationship] (space) to[bend right=20] node[relationship label, right, pos=0.5] {\scriptsize isLocationOf} (element);
\draw[relationship] (meter) -- node[relationship label, above, pos=0.4] {\scriptsize hasLocation} (space);
\draw[relationship](zone) to[bend right=25] node[relationship label, above, pos=0.7] {\scriptsize hasPoint}(setpoint);

\node[new entity] (person) at (-5.5, -1) {};
\node[new entity] (lease) at (-3, -1) {};

\node[attribute box, below=0.15cm of person] {
  \underline{REC\_Person} \\[1pt]
};

\node[attribute box, below=0.15cm of lease] {
  \underline{REC\_Lease} \\[1pt]
};

\draw[relationship] (lease) -- node[relationship label, above] {\scriptsize leasee} (person);
\draw[relationship] (lease) -- node[relationship label, above] {\scriptsize leaseOf} (space);

\node[entity, label=right:{\small Entities for Example 1}] at (4.5, -2) {};
\node[unified entity, label=right:{\small Unified entities (Ex1 and Ex2)}] at (4.5, -2.5) {};
\node[new entity, label=right:{\small\textcolor{blue!70}{Additional Entities for Example 2}}] at (4.5, -3) {};
\end{tikzpicture}

\caption{Theory extension $C$ for Examples 1 and 2. 
\textbf{Example 1} (entities within dashed boundary): IFC$\leftrightarrow$BRICK integration 
for generating BRICK operational models from IFC design data. The unified \texttt{Equipment} 
entity merges IFC's \texttt{IfcDistributionElement} with BRICK's \texttt{Equipment}; the 
unified \texttt{Location} merges \texttt{IfcSpace} with BRICK's \texttt{Location}. 
Formula~2 copies \texttt{deviceId} from \texttt{PropertySet} through to BRICK's 
\texttt{Point.timeseriesId}.
\textbf{Example 2} (all entities): Three-way integration adding REC for property management. 
The unified \texttt{Location} now also incorporates REC's \texttt{Room}, enabling 
cross-ontology queries: BRICK's \texttt{Meter} and \texttt{SetPoint} connect to REC's 
\texttt{Lease} through the shared spatial reference. Crucially, only IFC$\leftrightarrow$BRICK 
and IFC$\leftrightarrow$REC mappings are specified; the BRICK$\leftrightarrow$REC relationship 
emerges automatically through categorical composition. (Attributes are only shown for the unified entities, otherwise they remain the same as in Fig.~\ref{fig:individual_schemas}.}

\label{fig:threewayschema}
\end{figure}

\begin{lstlisting}[style=cql, float, caption={Example 1 CQL Query returning \texttt{spaceName} and \texttt{spaceArea} from IFC and \texttt{timeseriesId} from BRICK through the unified entity \texttt{Equipment}, reproducing the query from~\cite{mavrokapnidis_linked-data_2021}}, label={lst:mavro_query}]
query q = simple : Combined {
	from e:Equipment
	      
	attributes 
        IFC_spaceName -> e.hasLocation.spaceName
		IFC_spaceArea -> e.hasLocation.spaceArea
		BRICK_timeseriesId -> e.hasPoint.timeseriesId
}

\end{lstlisting}

\begin{table}[tbh]
\centering
\caption{Example 1 query result combining results from both IFC and BRICK, replicating the equivalent result from~\cite{mavrokapnidis_linked-data_2021}}
\begin{tabular}{lll}
\toprule
\texttt{IFC\textunderscore{}spaceName}& \texttt{IFC\textunderscore{}spaceArea} (m²) & \texttt{BRICK\textunderscore{}timeseriesId} \\
\midrule
Room 240 & 18.68 & TUC.245.77.R240 \\
Room 260 & 17.12 & TUC.245.77.R260 \\
Room 200 & 18.32 & TUC.245.77.R200 \\
Room 440 & 18.68 & TUC.245.77.R440 \\
Room 460 & 17.12 & TUC.245.77.R460\\
\bottomrule
\end{tabular}
\label{tab:mavro_results}
\end{table}

This example demonstrates favorable \emph{data complexity}: the integration rules are specified once and apply automatically to all 5 rooms (or 50, or 500) without additional effort. This contrasts with the approach in~\cite{mavrokapnidis_linked-data_2021}, where each \texttt{deviceId} must be manually extracted and matched to a manually created BRICK Point's \texttt{timeseriesId}, e.g., an effort that scales linearly with building size.

Beyond favorable data complexity, the categorical framework provides favorable \emph{query complexity} (\emph{compositional integration}): 
When integrating multiple ontologies, traditional approaches require explicit mappings between each pair, i.e., $O(n^2)$ specifications for $n$ ontologies. Category theory provides a principled alternative: mappings compose automatically (via logical entailment/deduction), so integrating three ontologies requires only two explicit specifications. We show this in the next example.

\subsection{Example 2: Three-Way Integration of IFC/BRICK with RealEstateCore}
\label{ss:example_three_way}
We now demonstrate this compositional scalability advantage by extending the previous scenario. We add RealEstateCore (REC) for property management~\cite{hammar_realestatecore_2019}, enabling tenant-based energy billing queries that span design, operations, and property management data. We note that there are harmonization efforts between BRICK and REC\footnote{\url{https://www.realestatecore.io/brickrec/}}. Here, our intention is to demonstrate that with the categorical approach, we can achieve data integration and data exchange for free: we will intentionally only specify mappings IFC$\leftrightarrow$BRICK and IFC$\leftrightarrow$REC and show that then REC$\leftrightarrow$BRICK is inferred automatically.

\begin{figure}
    \centering
    \includegraphics[width=\textwidth]{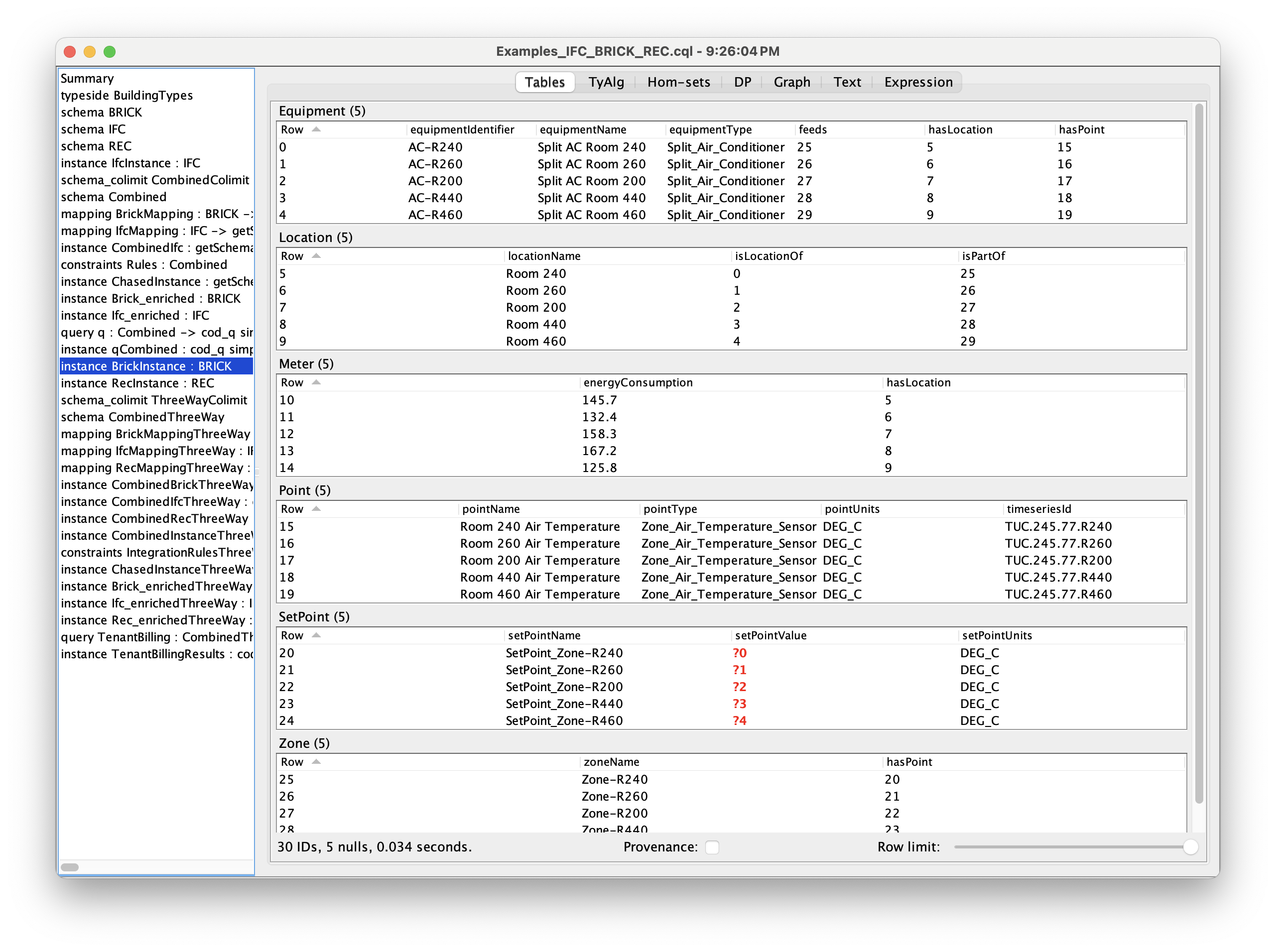}
    \caption{BRICK instance for Example 2 in CQL. Note the empty entry \texttt{setPointValue} for the \texttt{SetPoint} entity.}
    \label{fig:ex2brickInstance}
\end{figure}

\begin{figure}
    \centering
\includegraphics[width=\textwidth]{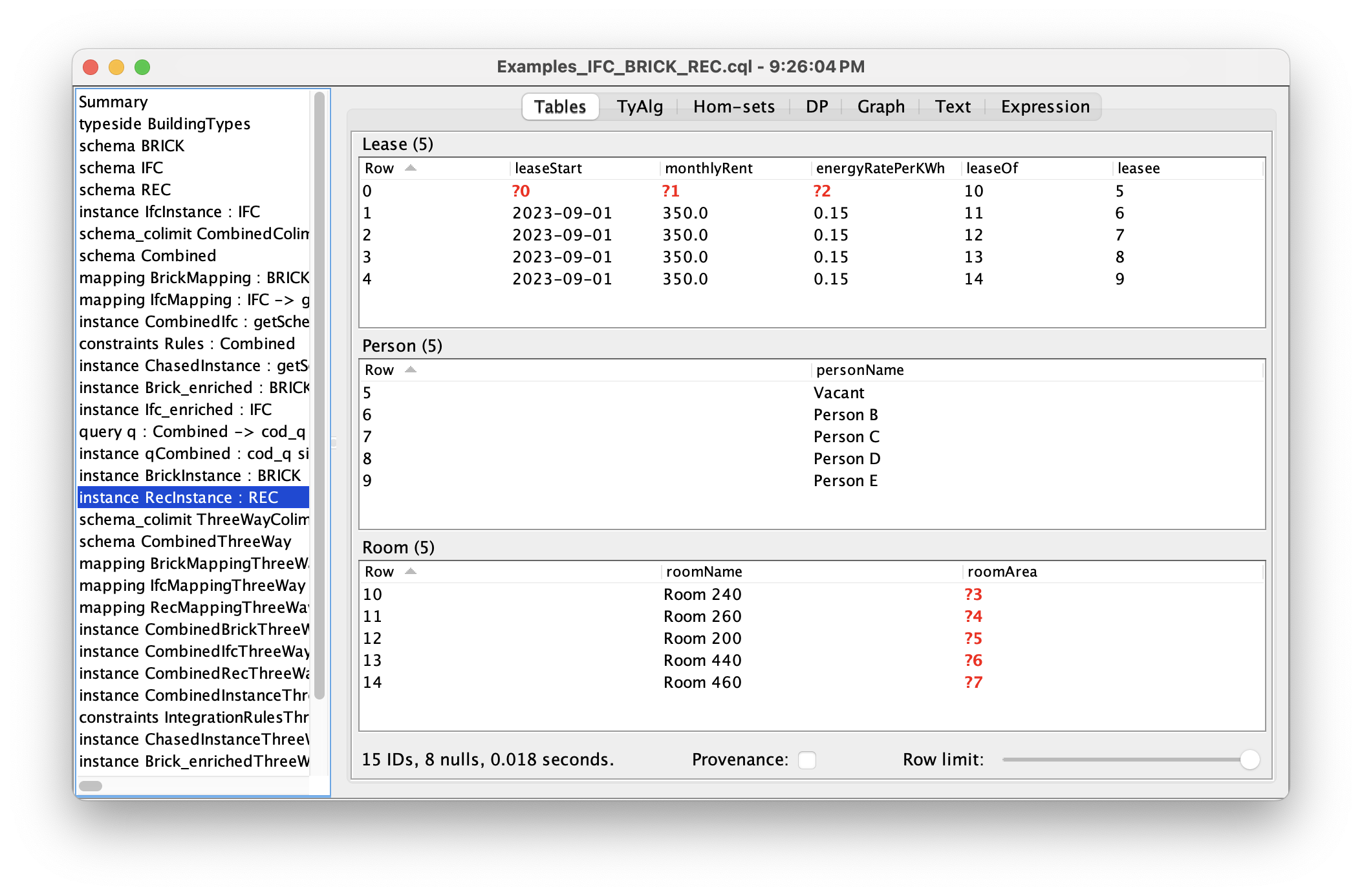}
    \caption{REC instance for Example 2 in CQL. Note the empty entries for \texttt{roomArea} for the \texttt{Room} entity. The top row in \texttt{Lease} is empty corresponding to the unleased room.}
    \label{fig:ex2recInstance}
\end{figure}

For the practical part of the example, we assume an existing BRICK instance (as opposed to Ex~1, where we created one from IFC). In particular, we associate an energy meter with each room that holds the last months aggregated energy use for billing purposes (REC). We also associate a (cooling) set point with each BRICK HVAC zone and will use vacancy information in REC to increase the set point in case of vacancy in order to reduce energy use. Finally, the REC instance contains lease information linked to each room. Crucially, \texttt{roomArea} is not initially available in REC and will be obtained later from IFC. The IFC instance is the same as previously (Fig~\ref{fig:ex1IfcInstance}), the BRICK and REC instances are shown in Figs~\ref{fig:ex2brickInstance} and~\ref{fig:ex2recInstance}.

Thus, the task is to combine IFC for room sizes, BRICK for energy meter readings and equipment associations, and REC for tenant assignments and billing information. We extend the theory extension from Example~1 to relate REC entities and additional BRICK components (see Fig.~\ref{fig:threewayschema}, blue entities):

\textbf{Entity Identifications}:
\begin{align*}
\texttt{Equipment} &\cong \texttt{IfcDistributionElement} \\
\texttt{Location} &\cong \texttt{IfcSpace}\\
\texttt{Room} &\cong \texttt{IfcSpace}
\end{align*}
The first two equations align BRICK with IFC entities. The third equation aligns IFC with REC. In particular, we do \emph{not} specify a direct BRICK$\leftrightarrow$REC relationship. The categorical construction automatically infers:
\[
\texttt{BRICK.Location} = \texttt{IFC.IfcSpace} = \texttt{REC.Room}
\]

This construction enables operations that span all three ontologies through a single unified Location. 

\textbf{Formulae 1 and 2} We can now identify existing entries based on structural relationships: Locations with identical \texttt{locationName} are unified. Formally:
\begin{align*}
\forall l_1, l_2:\texttt{Location},\ &l_1.\texttt{spaceName} = l_2.\texttt{roomName} \Rightarrow l_1 = l_2 \\
\forall l_1, l_2:\texttt{Location},\ &l_1.\texttt{spaceName} = l_2.\texttt{locationName} \Rightarrow l_1 = l_2
\end{align*}
Note that here we use equality of names for simplicity, in our experiments we have also used fuzzy equality allowing for slightly different room namings in each ontology.

\textbf{Formula 3} In this example, the REC instance does not contain the area of the rooms. But because we have aligned the \texttt{Location} entity in each ontology, and identified the same entities using Formulae 1 and 2, now we can easily exchange this data from IFC to REC. Formally:
\begin{align*}
    \forall l:\texttt{Location}, \texttt{l.roomArea = l.spaceArea} 
\end{align*}

\textbf{Formulae 4 and 5} We derive HVAC setpoints (BRICK) from occupancy status (REC):
\begin{align*}
\forall &l:\texttt{Lease},\ \texttt{levenshtein}(l.\texttt{leasee.personName}, \text{``Vacant''}) > 0 \\ &\Rightarrow l.\texttt{leaseOf.isPartOf.hasPoint.setPointValue} = 22 \\
\forall &l:\texttt{Lease},\ l.\texttt{leasee.personName} = \text{``Vacant''} \\ &\Rightarrow l.\texttt{leaseOf.isPartOf.hasPoint.setPointValue} = 26
\end{align*}
The first formula sets the comfort setpoint (22°C) for occupied rooms; the second sets the energy-saving setpoint (26°C) for vacant rooms. These rules traverse from REC's \texttt{Lease} through the unified \texttt{Location} to BRICK's \texttt{Zone} and its \texttt{SetPoint}, demonstrating how property management data can directly influence building operations through the categorical integration.

\textbf{Formula 6} Finally, as in the previous example, we align BRICK's and IFC's spatial relationship: 
\begin{align*}
\texttt{Equipment.hasLocation} &= \texttt{Equipment.elementInSpace} 
\end{align*}

\begin{figure}
    \centering
    \includegraphics[width=\textwidth]{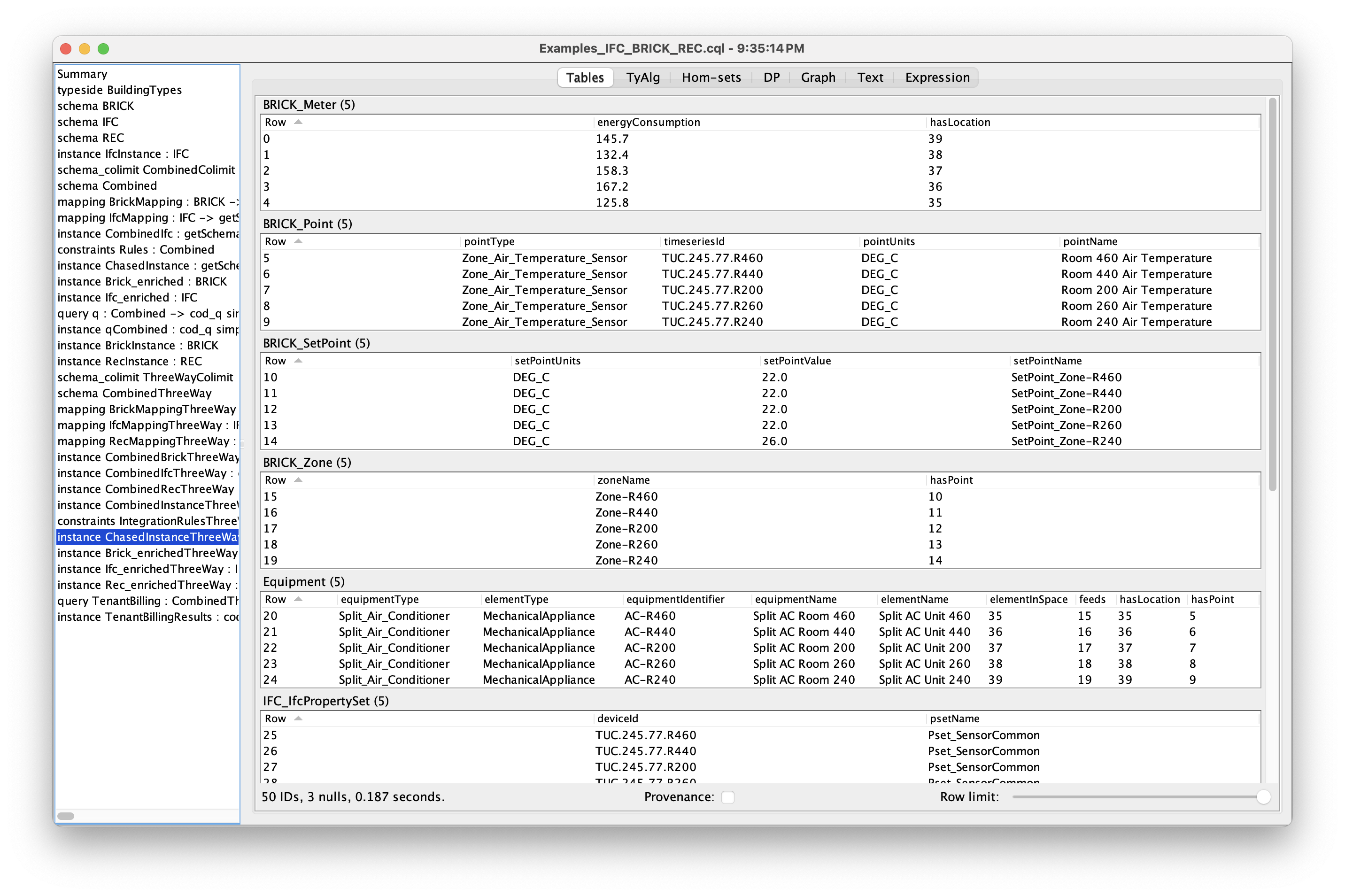}
    \caption{The Integrated Instance (1/2) of Example 2 after all rules have been applied. (cont'd in Fig.~\ref{fig:Ex2Chase2})}
    \label{fig:Ex2Chase1}
\end{figure}

\begin{figure}
    \centering
    \includegraphics[width=\textwidth]{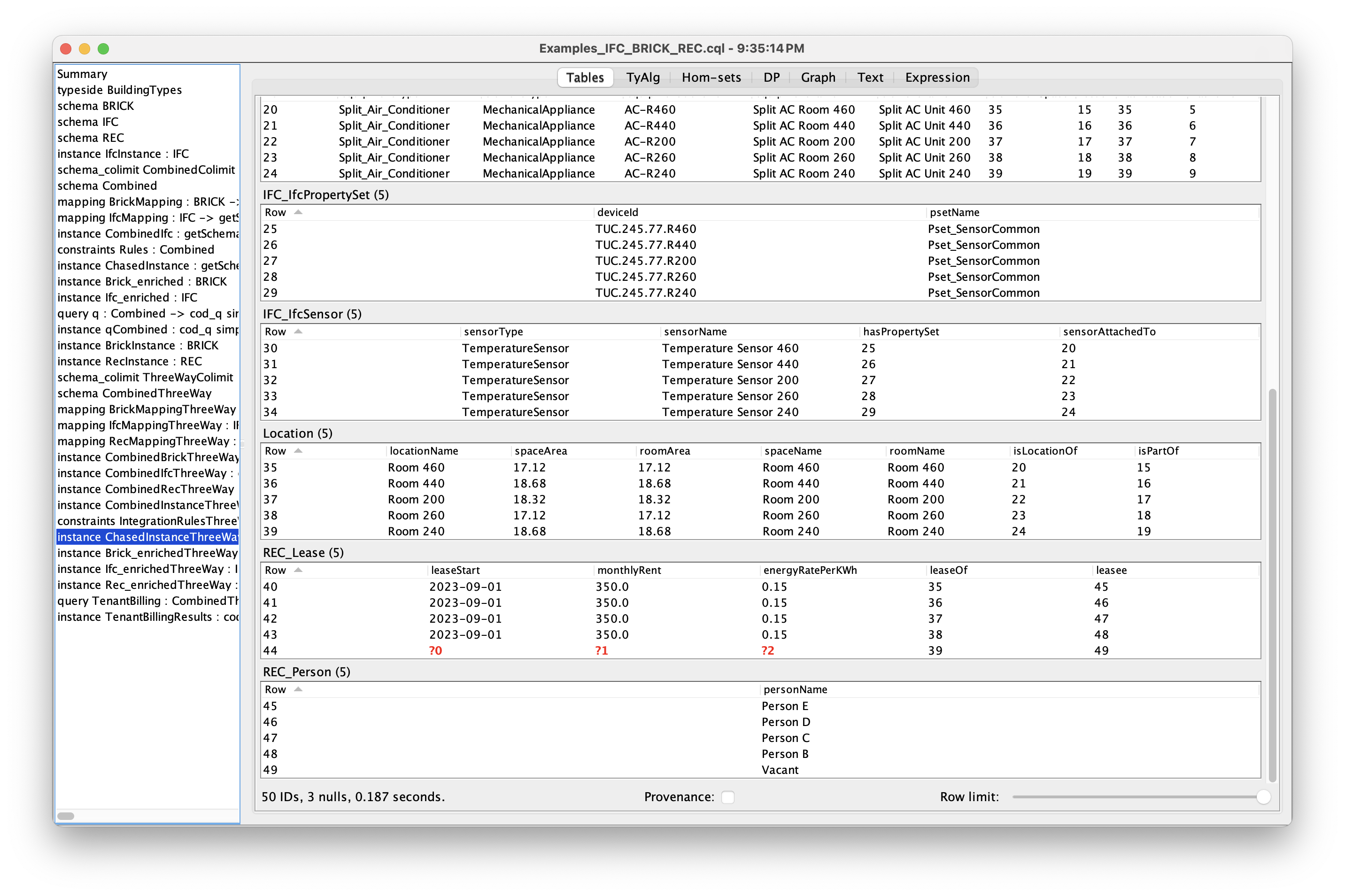}
    \caption{(cont'd from~\ref{fig:Ex2Chase1}) The Integrated Instance (2/2) of Example 2 after all rules have been applied.}
    \label{fig:Ex2Chase2}
\end{figure}

Figures~\ref{fig:Ex2Chase1} and~\ref{fig:Ex2Chase2} show the integrated instance after all rules have been applied. Notice how the unified entities \texttt{Equipment} and \texttt{Location} now contain all attributes from the individual instances. The query in Listing~\ref{lst:tenant_query} demonstrates the power of categorical composition: 
\texttt{lease.leaseOf} navigates from REC's \texttt{Lease} to the unified \texttt{Location}, 
which is simultaneously IFC's \texttt{IfcSpace}, BRICK's \texttt{Location}, and REC's \texttt{Room}. 
From there, \texttt{isLocationOf} reaches BRICK's \texttt{Equipment}, and the \texttt{where} 
clause joins with \texttt{BRICK\_Meter} through the same unified spatial reference. 
Table~\ref{tab:tenant_billing_results} shows the query results successfully combining data from REC, and BRICK. Note how \texttt{zoneSetPoint} depends on the vacancy (as mimicked by \texttt{personName}), and how \texttt{roomArea} has been obtained in REC after it has been copied there from IFC (via Formula 3).

\begin{lstlisting}[style=cql, float, caption={CQL query for tenant energy billing. The query joins REC lease data with BRICK meter data through the unified Location entity, demonstrating cross-ontology traversal without explicit BRICK$\leftrightarrow$REC mapping.}, label={lst:tenant_query}]
# Final Query (Examples of REC>BRICK through IFC)
query TenantBilling = simple : CombinedThreeWay {
    from lease:REC_Lease meter:BRICK_Meter
    
    where lease.leaseOf = meter.hasLocation
    
    attributes
        REC_personName -> lease.leasee.personName
        REC_roomName -> lease.leaseOf.roomName
        REC_roomArea -> lease.leaseOf.roomArea
        REC_monthlyRent -> lease.monthlyRent
        # Show Current SetPoints (REC > BRICK)
        BRICK_zoneSetPoint -> lease.leaseOf.isPartOf.hasPoint.setPointValue
        # Show Equipment (REC > BRICK)
        BRICK_Equipment -> lease.leaseOf.isLocationOf.equipmentName
        # Get energy consumption from BRICK points
        BRICK_energyUsed -> meter.energyConsumption       
 }
\end{lstlisting}

\begin{table}[tbh]
\centering
\caption{Query results for tenant energy billing. The BRICK$\leftrightarrow$REC join is achieved through the 
unified \texttt{Location} entity without explicit mapping between the two ontologies. Note that  \texttt{spaceArea} is queried from REC (after being copied from IFC), and that BRICK's \texttt{zoneSetPoint} depends on REC's \texttt{personName} for vacancy indication. }
\label{tab:tenant_billing_results}
\begin{tabular}{llccccc}
\toprule
\texttt{personName} & \texttt{roomName} & \texttt{roomArea} & \texttt{monthlyRent} & \texttt{zoneSetPoint} & \texttt{equipmentName} & \texttt{energyUsed} \\
(REC) & (REC) & (REC, m²) & (REC, €) & (BRICK, $^\circ$C) & (BRICK) & (BRICK, kWh) \\
\midrule
Vacant & Room 240 & 18.68 & - & 26.0 & Split AC Room 240 & 145.7 \\
Person B & Room 260 & 17.12 & 350.00 & 22.0 & Split AC Room 260 & 132.4 \\
Person C & Room 200 & 18.32 & 350.00 & 22.0 & Split AC Room 200 & 158.3 \\
Person D & Room 440 & 18.68 & 350.00 & 22.0 & Split AC Room 440 & 167.2 \\
Person E & Room 460 & 17.12 & 350.00 & 22.0 & Split AC Room 460 & 125.8 \\
\bottomrule
\end{tabular}
\end{table}

These two examples demonstrate the core capabilities of categorical data integration: Example 1 showed how theory extensions enable automated data generation from IFC to BRICK at commissioning, while Example 2 showed how compositional integration enables cross-ontology queries without explicit pairwise mappings. We now turn to discussing the broader implications of this framework, its advantages over current approaches, and the challenges that remain before production deployment.

\section{Discussion and Future Directions}
\label{s:discussion}

\subsection{Advantages Over Current Approaches}
Our categorical framework addresses fundamental limitations in current building data integration approaches through several key advantages: First, category theory provides mathematical structures specifically designed for structure-preserving transformations of sophisticated knowledge bases such as ontologies. This is the key paradigm shift. Solutions to categorical constructions such as the ``lifting problems'' of this paper preserve relationships by construction, rather than by post-hoc empirical testing (which can only prove the absence of errors, not correctness). 

Second,  categorical approaches require $O(n)$ formulae between $n$ sources to exchange data, as opposed to point-to-point mappings requiring $O(n^2)$ specifications, which also usually lack guarantees that the order point-to-point mappings are followed doesn't matter.  A special case of this is when an existing ontology is updated requiring reconstruction of the integration. Categorical approaches need to only extend the theory extension to recompute the data migrations, and order of integration doesn't matter. 

Third, integration rules in the theory extension are specified once and apply automatically to all instances regardless of database size. Traditional approaches require proportional effort: each entity pair must be individually matched and verified. These properties are key for scalability, practicability and integration across ever newly developed ontologies. 

Fourth, categorical integration enables \emph{cross-ontology data exchange}, i.e., deriving values in one ontology from data in another. Example~2 demonstrates this: HVAC setpoints in BRICK are derived from vacancy status in REC, traversing through the unified Location entity. The categorical framework makes such cross-domain policies declarative and verifiable. 

While we have not addressed this in detail in this paper, note that CQL operates independently of whether source data uses EXPRESS (IFC's native language), OWL/RDF (BRICK, BOT), or proprietary formats. 

\subsection{Implications for Building Digital Twins}
Digital twins require integrating heterogeneous data sources across the building lifecycle: design models (IFC), energy simulations (e.g., EnergyPlus), operational data (e.g., BRICK, BACnet), sensor streams (IoT platforms), occupant information, and many others especially if we also extend into urban scale digital twins. Current digital twin platforms typically address integration through proprietary data lakes, custom APIs, and manual data pipelines. As a result, integration costs and efforts are often prohibitive to deploy digital twins in practice at scale. 

We put forward our vision that the categorical framework provides the key foundations for managing this complexity, and, thus, is \textbf{the enabling technology for digital twins} (Fig.~\ref{fig:digital_twin_architecture}): Rather than maintaining pairwise mappings between ontologies, (which becomes unmanageable), or forcing all building data into a single reference ontology, (which becomes unwieldy, not to mention impossible if you want to relate the ontologies in more than one way, rather than one universal way), theory extensions allow modular composition. Each specialized ontology can be developed independently. New ontologies add $O(1)$ query complexity rather than requiring updates to all existing integrations. 

A practical illustration comes from recent efforts to extend BRICK for specialized domains, e.g., adding occupant behavior~\cite{luo_extending_2022} and energy flexibility metadata~\cite{li_semantic_2022}. In the reference ontology paradigm, e.g.,~\cite{fierro_shepherding_2020}, each such extension must be incorporated \emph{into} BRICK itself, progressively enlarging the reference schema. The categorical approach offers an alternative: occupant and flexibility ontologies remain independent first-order theories that compose with BRICK (and IFC, REC, etc.) through theory extensions, preserving modularity while enabling the same cross-ontology queries.

This modularity is the aforementioned key enabling feature for digital twins as they are inherently composed of heterogeneous data. Typically, if at all, one would manually maintain separate models with ad-hoc synchronization. The categorical framework provides principled model aggregation and disaggregation via data exchange.

\subsection{Limitations and Open Challenges}

Our framework has limitations that future research must address. First, theory extension specification currently requires knowledge beyond typical building science education. While CQL provides accessible syntax, defining the desired theory extension requires understanding the syntax and semantics of first-order logic. Tool development should target construction building professionals, not computer scientists, similar to how SQL users need not understand relational algebra's mathematical foundations. Visual editors for formulae specification, analogous to BIM authoring tools, would lower barriers to adoption. See~\cite{spivak_reglog_2019} for initial attempts in this direction.

Second, performance benchmarking against current SPARQL implementations remains future work, and empirical validation on realistic building models is needed. Comparisons should use standard benchmarks (e.g., IFC models from buildingSMART, BRICK models from published datasets) with metrics including query latency, throughput, and memory consumption.

Third, our proof-of-concept examples address only IFC-BRICK-REC integration. Extending to comprehensive building lifecycle coverage requires formalizing additional schemata: BuildingSync for energy audits, Project Haystack for tagging, proprietary BMS formats, etc. Each requires categorical formalization of existing schemata. Again, the compositional properties mean that integrating $n$ schemata requires $O(n)$ formulae specifications rather than $O(n^2)$ pairwise mappings, which is a fundamental scalability advantage.

\subsection{Broader Implications}

The building industry has no shortage of ideas for smart buildings, autonomous controls, grid-interactive communities, and AI-driven optimization. What it lacks is the computational infrastructure to reliably develop and deploy these innovations. Today, each new building algorithm requires custom integration code, manual data transformation, and extensive testing because there are no formal interfaces specifying how components should connect. Category theory provides exactly this missing infrastructure: \textbf{a mathematical foundation for building computation}. Just as smartphone operating systems enabled app ecosystems by providing formal APIs, categorical frameworks can enable building innovation ecosystems where researchers and practitioners develop algorithms with confidence they will integrate correctly.

Building decarbonization is inherently interdisciplinary: buildings account for approximately 30\% of global greenhouse gas emissions, yet achieving deep reductions requires coordinating multiple technical domains that often work independently. Effective decarbonization must simultaneously address amongst others, urban development patterns, grid decarbonization, technology adoption, and energy efficiency; each with specialized knowledge and methods~\cite{felkner_impact_2024}. Forcing these into a single unified ontology creates unmanageable complexity. Category theory enables a different approach supporting genuine interdisciplinary collaboration: each domain develops models in their own native formalisms while categorical methods ensure consistency at integration boundaries. See, e.g., ~\cite{nolan_compositional_2020} for an example from power systems and distributed energy management.

\section{Conclusion}
\label{s:conclusion}

Building data integration requires mathematical foundations that current ontology-based approaches lack. We have demonstrated how category theory provides these foundations through proof-of-concept implementations integrating IFC design data with BRICK operational models and REC property management database. We formalize ontologies as theories and use theory extensions and lifting problems to achieve provable structure preservation correctness properties (the result of our data integration / exchange is guaranteed by construction to satisfy the formulae that define it), automated bidirectional migration, compile-time error detection, and data-independent rules that apply uniformly regardless of building size (rooms, equipment).

As buildings transition toward digital twins requiring real-time integration of design, operational, and simulation data from an increasing number of ontologies, mathematical foundations become essential for managing complexity at scale. Category theory provides these foundations, transforming data integration from ad-hoc schema mapping into rigorous mathematical theory with provable properties. This framework will enable next-generation building digital twins and data platforms that evolve with technological requirements while maintaining consistency guarantees impossible with current approaches.

\section*{Acknowledgement}
\label{s:acknowledgement}
The authors would like to thank David Spivak for inspiring discussions.

\bibliographystyle{unsrt}
\bibliography{references, references-2}

\newpage
\appendix
\section{CQL Listing for Example 1 and 2}
\label{app:cql_code}
\begin{lstlisting}[style=cql]
# Tested with java v 21.0.8 (12.250)

typeside BuildingTypes = literal {
    external_types
        String -> "java.lang.String"
        UUID -> "java.util.UUID"
        Double -> "java.lang.Double"
        Boolean -> "java.lang.Boolean"
        Integer -> "java.lang.Integer"
    external_parsers
        String -> "x => x"
        UUID -> "x => java.util.UUID.fromString(x)"
        Double -> "x => java.lang.Double.parseDouble(x)"
        Boolean -> "x => java.lang.Boolean.parseBoolean(x)"
        Integer -> "x => java.lang.Integer.parseInt(x)" 
    external_functions
    	"<" : Integer, Integer -> Boolean = "(a,b) => a < b"
    	">" : Integer, Integer -> Boolean = "(a,b) => a > b"
        levenshtein : String, String -> Integer = "(a,b) => org.apache.commons.text.similarity.LevenshteinDistance.getDefaultInstance().apply(a,b)"
}

# BRICK Schema - Operation
schema BRICK = literal : BuildingTypes {
    entities
        Equipment
        Point
        Location
        Zone
        Meter
        SetPoint
    foreign_keys
        hasPoint : Equipment -> Point
        hasLocation : Equipment -> Location
        isLocationOf : Location -> Equipment
        feeds : Equipment -> Zone
        isPartOf : Location -> Zone
        hasLocation : Meter -> Location
        hasPoint : Zone -> SetPoint
    path_equations
    	Equipment = Equipment.hasLocation.isLocationOf # inverse definitions
		Location = Location.isLocationOf.hasLocation   # inverse definitions
    attributes
        equipmentIdentifier : Equipment -> String
        equipmentName : Equipment -> String
        equipmentType : Equipment -> String
        pointName : Point -> String
        pointType : Point -> String
        pointUnits : Point -> String
        timeseriesId : Point -> String # BMS identifier for time-series data
        energyConsumption : Meter -> Double # kWh for billing
        locationName : Location -> String
        zoneName : Zone -> String
		setPointName : SetPoint -> String
		setPointValue : SetPoint -> Double
		setPointUnits : SetPoint -> String
}

# IFC Schema - Design
schema IFC = literal : BuildingTypes {
    entities
        IfcSpace
        IfcDistributionElement
        IfcSensor
        IfcPropertySet
    foreign_keys
        elementInSpace : IfcDistributionElement -> IfcSpace
        sensorAttachedTo : IfcSensor -> IfcDistributionElement
        hasPropertySet : IfcSensor -> IfcPropertySet
    attributes
        spaceName : IfcSpace -> String
        spaceArea : IfcSpace -> Double
        elementName : IfcDistributionElement -> String
        elementType : IfcDistributionElement -> String
        sensorName : IfcSensor -> String
        sensorType : IfcSensor -> String
        psetName : IfcPropertySet -> String
        deviceId : IfcPropertySet -> String 
}

# RealEstateCore Schema - Property management
schema REC = literal : BuildingTypes {
    entities
        Room 
        Person
        Lease
    foreign_keys
    	leaseOf : Lease -> Room
        leasee : Lease -> Person
    attributes
        roomName : Room -> String
        roomArea : Room -> Double  
        personName : Person -> String
        leaseStart : Lease -> String
        monthlyRent : Lease -> Double
        energyRatePerKWh : Lease -> Double 
}

###########################################################################
# EXAMPLE 1: GENERATE BRICK FROM IFC (RECREATE MAVROKAPNIDIS'21), NO REC  #
###########################################################################

# IFC Instance - Design/construction data from BIM model
instance IfcInstance = literal : IFC {
    generators
        # Spaces - 5 rooms
        space_240 space_260 space_200 space_440 space_460 : IfcSpace   
        # Distribution Elements - 5 AC units
        ac_elem_240 ac_elem_260 ac_elem_200 ac_elem_440 ac_elem_460 : IfcDistributionElement       
        # Sensors - 5 temperature sensors
        sensor_240 sensor_260 sensor_200 sensor_440 sensor_460 : IfcSensor       
        # Property Sets
        pset_240 pset_260 pset_200 pset_440 pset_460 : IfcPropertySet
        
    equations
        # Space 240
        spaceName(space_240) = "Room 240"
        spaceArea(space_240) = "18.68"
        elementName(ac_elem_240) = "Split AC Unit 240"
        elementType(ac_elem_240) = "MechanicalAppliance"
        elementInSpace(ac_elem_240) = space_240
        sensorName(sensor_240) = "Temperature Sensor 240"
        sensorType(sensor_240) = "TemperatureSensor"
        sensorAttachedTo(sensor_240) = ac_elem_240
        hasPropertySet(sensor_240) = pset_240
        psetName(pset_240) = "Pset_SensorCommon"
        deviceId(pset_240) = "TUC.245.77.R240" 
        
        # Space 260
        spaceName(space_260) = "Room 260"
        spaceArea(space_260) = "17.12"
        elementName(ac_elem_260) = "Split AC Unit 260"
        elementType(ac_elem_260) = "MechanicalAppliance"
        elementInSpace(ac_elem_260) = space_260
        sensorName(sensor_260) = "Temperature Sensor 260"
        sensorType(sensor_260) = "TemperatureSensor"
        sensorAttachedTo(sensor_260) = ac_elem_260
        hasPropertySet(sensor_260) = pset_260
        psetName(pset_260) = "Pset_SensorCommon"
        deviceId(pset_260) = "TUC.245.77.R260"
        
        # Space 200
        spaceName(space_200) = "Room 200"
        spaceArea(space_200) = "18.32"
        elementName(ac_elem_200) = "Split AC Unit 200"
        elementType(ac_elem_200) = "MechanicalAppliance"
        elementInSpace(ac_elem_200) = space_200
        sensorName(sensor_200) = "Temperature Sensor 200"
        sensorType(sensor_200) = "TemperatureSensor"
        sensorAttachedTo(sensor_200) = ac_elem_200
        hasPropertySet(sensor_200) = pset_200
        psetName(pset_200) = "Pset_SensorCommon"
        deviceId(pset_200) = "TUC.245.77.R200"
        
        # Space 440
        spaceName(space_440) = "Room 440"
        spaceArea(space_440) = "18.68"
        elementName(ac_elem_440) = "Split AC Unit 440"
        elementType(ac_elem_440) = "MechanicalAppliance"
        elementInSpace(ac_elem_440) = space_440
        sensorName(sensor_440) = "Temperature Sensor 440"
        sensorType(sensor_440) = "TemperatureSensor"
        sensorAttachedTo(sensor_440) = ac_elem_440
        hasPropertySet(sensor_440) = pset_440
        psetName(pset_440) = "Pset_SensorCommon"
        deviceId(pset_440) = "TUC.245.77.R440"
        
        # Space 460
        spaceName(space_460) = "Room 460"
        spaceArea(space_460) = "17.12"
        elementName(ac_elem_460) = "Split AC Unit 460"
        elementType(ac_elem_460) = "MechanicalAppliance"
        elementInSpace(ac_elem_460) = space_460
        sensorName(sensor_460) = "Temperature Sensor 460"
        sensorType(sensor_460) = "TemperatureSensor"
        sensorAttachedTo(sensor_460) = ac_elem_460
        hasPropertySet(sensor_460) = pset_460
        psetName(pset_460) = "Pset_SensorCommon"
        deviceId(pset_460) = "TUC.245.77.R460"
}

# This specifies HOW IFC and BRICK schemas should be unified
schema_colimit CombinedColimit = quotient BRICK + IFC : BuildingTypes {
    entity_equations
        # Equipment in BRICK = Distribution elements in IFC
        BRICK.Equipment = IFC.IfcDistributionElement      
        # Locations in BRICK = Spaces in IFC
        BRICK.Location = IFC.IfcSpace		
    path_equations
        # Equipment location relationships align
        Equipment.hasLocation = Equipment.elementInSpace
}

# Extract the combined schema and inclusion mappings
schema Combined = getSchema CombinedColimit
mapping BrickMapping = getMapping CombinedColimit BRICK
mapping IfcMapping = getMapping CombinedColimit IFC

# SIGMA: Push data forward into combined schema
instance CombinedIfc = sigma IfcMapping IfcInstance

# Constraints/Rules - Additional logic to complete the integration
constraints Rules = literal : Combined { 
	# Copy deviceId from IFC PropertySet into BRICK timeseriesId
	forall s:IFC_IfcSensor p:BRICK_Point
	where p = s.sensorAttachedTo.hasPoint
	-> where p.timeseriesId = s.hasPropertySet.deviceId	
}

# Integrated Instance with Rules fullfilled
instance ChasedInstance = chase Rules CombinedIfc

# DELTA: Pull enriched data back to original schemas
instance Brick_enriched = delta BrickMapping ChasedInstance
instance Ifc_enriched = delta IfcMapping ChasedInstance

# Query equivalent to Mavrokapnidis'21
query q = simple : Combined {
	from e:Equipment
	      
	attributes IFC_spaceName -> e.hasLocation.spaceName
			   IFC_spaceArea -> e.hasLocation.spaceArea
			   BRICK_timeseriesId -> e.hasPoint.timeseriesId
}
instance qCombined = eval q ChasedInstance


#############################################################################################
# EXAMPLE 2: REC > IFC > BRICK Link property management (REC) to operations (BRICK) through #
# Design (IFC) 																				#
#############################################################################################

# Reuse the above schema and instances, but now create a BRICK instance assume that it has been done during commisioning
instance BrickInstance = literal : BRICK {
    generators
        # Equipment - 5 AC units
        ac_R240 ac_R260 ac_R200 ac_R440 ac_R460: Equipment       
        # Points - Temperature sensors 
        temp_R240 temp_R260 temp_R200 temp_R440 temp_R460 : Point       
        # Locations - Physical rooms
        room_240 room_260 room_200 room_440 room_460 : Location      
        # Zones - HVAC zones (each room is its own zone)
        zone_240 zone_260 zone_200 zone_440 zone_460 : Zone
        # Meters - each meter meters one HVAC Zone
        em_R240 em_R260 em_R200 em_R440 em_R460 : Meter
        # SetPoints - each zone has a SetPoint
        stp_z240 stp_z260 stp_z200 stp_z440 stp_z460: SetPoint
        
    equations
        # AC Unit R240 (Room 240, Floor 2)
        equipmentIdentifier(ac_R240) = "AC-R240"
        equipmentName(ac_R240) = "Split AC Room 240"
        equipmentType(ac_R240) = "Split_Air_Conditioner"
        hasLocation(ac_R240) = room_240
        feeds(ac_R240) = zone_240
        hasPoint(ac_R240) = temp_R240

        # Temperature Point R240
        pointName(temp_R240) = "Room 240 Air Temperature"
        pointType(temp_R240) = "Zone_Air_Temperature_Sensor"
        pointUnits(temp_R240) = "DEG_C"
        timeseriesId(temp_R240) = "TUC.245.77.R240"
        energyConsumption(em_R240) = "145.7"
        hasLocation(em_R240) = room_240
        
        # Location R240
        locationName(room_240) = "Room 240"
        isPartOf(room_240) = zone_240
        
        # Zone R240
        zoneName(zone_240) = "Zone-R240"
        hasPoint(zone_240) = stp_z240
		setPointName(stp_z240) = "SetPoint_Zone-R240"
		setPointUnits(stp_z240) = "DEG_C"
      
        # AC Unit R260
        equipmentIdentifier(ac_R260) = "AC-R260"
        equipmentName(ac_R260) = "Split AC Room 260"
        equipmentType(ac_R260) = "Split_Air_Conditioner"
        hasLocation(ac_R260) = room_260
        feeds(ac_R260) = zone_260
        hasPoint(ac_R260) = temp_R260

        pointName(temp_R260) = "Room 260 Air Temperature"
        pointType(temp_R260) = "Zone_Air_Temperature_Sensor"
        pointUnits(temp_R260) = "DEG_C"
        timeseriesId(temp_R260) = "TUC.245.77.R260"
		energyConsumption(em_R260) = "132.4"
        hasLocation(em_R260) = room_260
        
        locationName(room_260) = "Room 260"
        isPartOf(room_260) = zone_260
        zoneName(zone_260) = "Zone-R260"
		hasPoint(zone_260) = stp_z260
		setPointName(stp_z260) = "SetPoint_Zone-R260"
		setPointUnits(stp_z260) = "DEG_C"

        
        # AC Unit R200
        equipmentIdentifier(ac_R200) = "AC-R200"
        equipmentName(ac_R200) = "Split AC Room 200"
        equipmentType(ac_R200) = "Split_Air_Conditioner"
        hasLocation(ac_R200) = room_200
        feeds(ac_R200) = zone_200
        hasPoint(ac_R200) = temp_R200

        pointName(temp_R200) = "Room 200 Air Temperature"
        pointType(temp_R200) = "Zone_Air_Temperature_Sensor"
        pointUnits(temp_R200) = "DEG_C"
        timeseriesId(temp_R200) = "TUC.245.77.R200"
        energyConsumption(em_R200) = "158.3"
        hasLocation(em_R200) = room_200
        
        locationName(room_200) = "Room 200"
        isPartOf(room_200) = zone_200
        zoneName(zone_200) = "Zone-R200"
        hasPoint(zone_200) = stp_z200
		setPointName(stp_z200) = "SetPoint_Zone-R200"
		setPointUnits(stp_z200) = "DEG_C"
      
        # AC Unit R440
        equipmentIdentifier(ac_R440) = "AC-R440"
        equipmentName(ac_R440) = "Split AC Room 440"
        equipmentType(ac_R440) = "Split_Air_Conditioner"
        hasLocation(ac_R440) = room_440
        feeds(ac_R440) = zone_440
        hasPoint(ac_R440) = temp_R440
        
        pointName(temp_R440) = "Room 440 Air Temperature"
        pointType(temp_R440) = "Zone_Air_Temperature_Sensor"
        pointUnits(temp_R440) = "DEG_C"
        timeseriesId(temp_R440) = "TUC.245.77.R440"
        energyConsumption(em_R440) = "167.2"
        hasLocation(em_R440) = room_440
        
        locationName(room_440) = "Room 440"
        isPartOf(room_440) = zone_440
        zoneName(zone_440) = "Zone-R440"
        hasPoint(zone_440) = stp_z440
		setPointName(stp_z440) = "SetPoint_Zone-R440"
		setPointUnits(stp_z440) = "DEG_C"
		
        # AC Unit R460
        equipmentIdentifier(ac_R460) = "AC-R460"
        equipmentName(ac_R460) = "Split AC Room 460"
        equipmentType(ac_R460) = "Split_Air_Conditioner"
        hasLocation(ac_R460) = room_460
        feeds(ac_R460) = zone_460
        hasPoint(ac_R460) = temp_R460
		
        pointName(temp_R460) = "Room 460 Air Temperature"
        pointType(temp_R460) = "Zone_Air_Temperature_Sensor"
        pointUnits(temp_R460) = "DEG_C"
        timeseriesId(temp_R460) = "TUC.245.77.R460"
        energyConsumption(em_R460) = "125.8"
        hasLocation(em_R460) = room_460
        
        locationName(room_460) = "Room 460"
        isPartOf(room_460) = zone_460
        zoneName(zone_460) = "Zone-R460"
        hasPoint(zone_460) = stp_z460
		setPointName(stp_z460) = "SetPoint_Zone-R460"
		setPointUnits(stp_z460) = "DEG_C"
}

# REC Instance - Property management data
instance RecInstance = literal : REC {
    generators   
        # Rooms that can be leased
        rec_room_240 rec_room_260 rec_room_200 rec_room_440 rec_room_460 : Room
        # Leasee
        person_A person_B person_C person_D person_E : Person
        # Leases
        lease_240 lease_260 lease_200 lease_440 lease_460 : Lease
        
    equations    
        # Room 240 - unleased
        roomName(rec_room_240) = "Room 240"
        personName(person_A) = "Vacant" # set name to vacant to mimic unrented room
        leaseOf(lease_240) = rec_room_240
        leasee(lease_240) = person_A
        
        # Room 260 - leased by B
        roomName(rec_room_260) = "Room 260"
        personName(person_B) = "Person B"
        leaseOf(lease_260) = rec_room_260
        leasee(lease_260) = person_B
        leaseStart(lease_260) = "2023-09-01"
        monthlyRent(lease_260) = "350.00"
        energyRatePerKWh(lease_260) = "0.15"
        
        # Room 200 - leased by C
        roomName(rec_room_200) = "Room 200"
        personName(person_C) = "Person C"
        leaseOf(lease_200) = rec_room_200
        leasee(lease_200) = person_C
        leaseStart(lease_200) = "2023-09-01"
        monthlyRent(lease_200) = "350.00"
        energyRatePerKWh(lease_200) = "0.15"
        
        # Room 440 - leased by D
        roomName(rec_room_440) = "Room 440"
        personName(person_D) = "Person D"
        leaseOf(lease_440) = rec_room_440
        leasee(lease_440) = person_D
        leaseStart(lease_440) = "2023-09-01"
        monthlyRent(lease_440) = "350.00"
        energyRatePerKWh(lease_440) = "0.15"
        
        # Room 460 - leased by E
        roomName(rec_room_460) = "Room 460"
        personName(person_E) = "Person E"
        leaseOf(lease_460) = rec_room_460
        leasee(lease_460) = person_E
        leaseStart(lease_460) = "2023-09-01"
        monthlyRent(lease_460) = "350.00"
        energyRatePerKWh(lease_460) = "0.15"
}

# THREE-WAY INTEGRATION: IFC + BRICK + REC
# This demonstrates O(n) integration: we only specify pairwise relationships
schema_colimit ThreeWayColimit = quotient BRICK + IFC + REC : BuildingTypes {
    entity_equations
        # IFC + BRICK
        BRICK.Equipment = IFC.IfcDistributionElement
        BRICK.Location = IFC.IfcSpace
		
        # REC + IFC 
        IFC.IfcSpace = REC.Room        
    path_equations
        # IFC + BRICK relationships
        Equipment.hasLocation = Equipment.elementInSpace
}

# Extract the combined schema and inclusion mappings
schema CombinedThreeWay = getSchema ThreeWayColimit
mapping BrickMappingThreeWay = getMapping ThreeWayColimit BRICK
mapping IfcMappingThreeWay = getMapping ThreeWayColimit IFC
mapping RecMappingThreeWay = getMapping ThreeWayColimit REC

# SIGMA: Push data forward into combined schema
instance CombinedBrickThreeWay = sigma BrickMappingThreeWay BrickInstance
instance CombinedIfcThreeWay = sigma IfcMappingThreeWay IfcInstance
instance CombinedRecThreeWay = sigma RecMappingThreeWay RecInstance

# Merge all three instances
instance CombinedInstanceThreeWay = coproduct CombinedBrickThreeWay + CombinedIfcThreeWay + CombinedRecThreeWay : CombinedThreeWay

# Constraints/Rules - The key to compositional integration!
constraints IntegrationRulesThreeWay = literal : CombinedThreeWay {

	# From Previous Example:
	# Copy deviceId from IFC PropertySet into BRICK timeseriesId
	forall s:IFC_IfcSensor p:BRICK_Point
	where p = s.sensorAttachedTo.hasPoint
	-> where p.timeseriesId = s.hasPropertySet.deviceId
	
    # Merge rows with the same locationName (could be fuzzyfied)
	forall l1 l2:Location where
	l1.spaceName = l2.roomName -> where l1 = l2

	forall l1 l2:Location where
	l1.spaceName = l2.locationName -> where l1 = l2

	# Data Exchange IFC -> REC
	# Copy IFC.spaceArea to REC.roomArea
	forall l:Location -> where l.roomArea = l.spaceArea

	# Data Exchange REC -> BRICK (through IFC)
	# Set BRICK Zone SetPoints to 22C if Room is occupied
	forall l:REC_Lease where (levenshtein(l.leasee.personName,"Vacant") ">" 0) = true
		-> where l.leaseOf.isPartOf.hasPoint.setPointValue = "22"
	# Set BRICK Zone SetPoints to 26C if Rooms has no tenant
	forall l:REC_Lease where l.leasee.personName = "Vacant"
		-> where l.leaseOf.isPartOf.hasPoint.setPointValue = "26"

}

# Integrated Instance with Rules fullfilled
instance ChasedInstanceThreeWay = chase IntegrationRulesThreeWay CombinedInstanceThreeWay { options require_consistency=false }

# DELTA: Pull enriched data back to original schemas
instance Brick_enrichedThreeWay = delta BrickMappingThreeWay ChasedInstanceThreeWay
instance Ifc_enrichedThreeWay = delta IfcMappingThreeWay ChasedInstanceThreeWay
instance Rec_enrichedThreeWay = delta RecMappingThreeWay ChasedInstanceThreeWay

# Final Query (Examples of REC>BRICK through IFC)
query TenantBilling = simple : CombinedThreeWay {
    from lease:REC_Lease meter:BRICK_Meter
    
    where lease.leaseOf = meter.hasLocation
    
    attributes
        REC_personName -> lease.leasee.personName
        REC_roomName -> lease.leaseOf.roomName
        REC_roomArea -> lease.leaseOf.roomArea
        REC_monthlyRent -> lease.monthlyRent
        # Show Current SetPoints (REC > BRICK)
        BRICK_zoneSetPoint -> lease.leaseOf.isPartOf.hasPoint.setPointValue
        # Show Equipment (REC > BRICK)
        BRICK_Equipment -> lease.leaseOf.isLocationOf.equipmentName
        # Get energy consumption from BRICK points
        BRICK_energyUsed -> meter.energyConsumption       
 }

instance TenantBillingResults = eval TenantBilling ChasedInstanceThreeWay
\end{lstlisting}

\end{document}